\documentclass[10pt,conference,compsocconf]{IEEEtran}

\usepackage{cite}
\usepackage[pdftex]{graphicx}
\usepackage[caption=false,font=footnotesize]{subfig}
\graphicspath{{./figures/}}
\DeclareGraphicsExtensions{.pdf}

\newcommand{\elem}[1]{{\me{#1}}}
\newcommand{\me}[1]{\textsf{\small #1}}
\usepackage{url}
\makeatletter
\def\ps@IEEEtitlepagestyle{
	\def\@oddfoot{\mycopyrightnotice \thepage}
	\def\@evenfoot{}
}
\def\mycopyrightnotice{
	{\footnotesize
		\begin{minipage}{\textwidth}
			\centering
			\textcopyright~2012 IEEE. Personal use of this material is permitted.
			Permission from IEEE must be obtained for all other uses, in any current or future media, including reprinting/republishing this material for advertising or promotional purposes, creating new collective works, for resale or redistribution to servers or lists, or reuse of any copyrighted component of this work in other works. 
			{\tt DOI:} \url{https://doi.org/10.1109/SEAMS.2012.6224399}
		\end{minipage}
	}
}
\usepackage[hidelinks]{hyperref}
\hypersetup{
	bookmarks=true,         
	unicode=false,          
	pdftoolbar=true,        
	pdfmenubar=true,        
	pdffitwindow=false,     
	pdftitle={A Language for Feedback Loops in Self-Adaptive Systems: Executable Runtime Megamodels},
	pdfauthor={Thomas Vogel and Holger Giese},
	pdfsubject={},
	pdfcreator={},
	pdfproducer={},
	pdfkeywords={self-adaptation; feedback loop; modeling language; megamodel; runtime models; model interpretation;},
	pdfnewwindow=true,
}

\begin{document}

\title{A Language for Feedback Loops in Self-Adaptive Systems:\\Executable Runtime Megamodels}

\author{\IEEEauthorblockN{Thomas Vogel and Holger Giese}
\IEEEauthorblockA{
Hasso Plattner Institute at the University of Potsdam, Germany\\
$\{$thomas.vogel$|$holger.giese$\}$@hpi.uni-potsdam.de}
}

\maketitle
\pagestyle{plain}
\begin{abstract}
The development of self-adaptive software requires the engineering of proper feedback loops where an adaptation logic controls the underlying software. The adaptation logic often describes the adaptation by using runtime models representing the underlying software and steps such as analysis and planning that operate on these runtime models. To systematically address this interplay, runtime megamodels, which are specific runtime models that have themselves runtime models as their elements and that also capture the relationships between multiple runtime models, have been proposed.

In this paper, we go one step further and present a modeling language for runtime megamodels that considerably eases the development of the adaptation logic by providing a domain-specific modeling approach and a runtime interpreter for this part of a self-adaptive system. This supports development by modeling the feedback loops explicitly and at a higher level of abstraction. Moreover, it permits to build complex solutions where multiple feedback loops interact or operate on top of each other, which is leveraged by keeping the megamodels explicit and alive at runtime and by interpreting them.
\end{abstract}

\section{Introduction}
\label{sec:introduction}

The development of self-adaptive software following the \emph{external approach}~\cite{Salehie&Tahvildari2009} separates the software into the domain logic and the adaptation logic. In between both, a \emph{feedback loop} ensures that the adaptation logic dynamically adjusts the domain logic in response to changing requirements and observed changes in the domain logic and its environment.

The separation eases the development of the domain logic because it decouples the domain logic from the adaptation logic, and both are integrated by well-defined sensor and effector interfaces. Thus, the adaptation logic is kept separate and does not directly contribute to the complexity of the domain logic. However, the feedback loop then becomes a crucial element of the overall software architecture, which has to be understood and explicitly designed (cf.~\cite{PezzeMS08,Brun+2009}).

Additionally, in more advanced scenarios even multiple feedback loops have to be considered. 
On the one hand, the adaptation logic may not necessarily employ only a single feedback loop but rather multiple of them in parallel to handle different concerns such as self-healing/repair or self-optimization~\cite{VogelNHGB10,VG10}.
On the other hand, there are also cases where the feedback loops have to operate on top of each other as, for example, needed for the different layers of the reference architecture for self-managed systems proposed in~\cite{Kramer&Magee2007} or in hierarchical structures with internal layers~\cite{Hestermeyer+2004}.

A specific approach to support the development of the adaptation logic is to leverage the benefits of \emph{model-driven engineering} (MDE) for representing the adaptable sub-system, like the domain logic, at runtime. Such representations serve as a basis for the adaptation logic's computations and they are realized by \emph{runtime models} built on MDE principles~\cite{MC.2009.326}. Other approaches, like~\cite{GarCHSS04,georgas-computer09}, also employ runtime representations, however, they are based on architecture description languages but not on MDE principles.

In this context, it is likely that the adaptation logic does not employ a single runtime model but rather multiple and specialized models at the same time to handle different concerns such as failures or performance with different feedback loops supporting self-healing or self-optimization, respectively. This makes it necessary to simultaneously consider multiple runtime models and the interplay between them when engineering and executing the adaptation logic.

In the MDE field of model management for model-driven software development, \emph{megamodels} refer to models that have models as their elements and that capture the relationships between multiple models in the form of model operations, like model transformations (cf.~\cite{BDDFB07,Bezivin_et_al:2003,Bezivin+2004,favre:DSP:2005:13}). Consequently, we proposed in~\cite{VogelSG11} to utilize similar concepts for self-adaptive software systems that are based on runtime models. By employing \emph{runtime megamodels}, the different runtime models, the interplay between these models, and the model operations performing adaptation steps and working on these models can be explicitly maintained at runtime and thus beyond the initial development-time of the system.

In this paper, we go one step further and present a complete modeling language for runtime megamodels that considerably eases the development of the adaptation logic by supporting a domain-specific modeling solution and a runtime interpreter for this part of a self-adaptive system.
The major benefits of the solution are the following:
(1) the feedback loops are explicitly described in the megamodels, 
(2) the adaptation is specified at a higher level of abstraction in the megamodels by a number of model operations that work on runtime models,
(3) the megamodels are kept alive at runtime and following an interpreter approach, the runtime models and the control flow between the model operations can be easily adapted at runtime, and
(4) composing and especially the stacking of feedback loops resp. megamodels becomes possible without any need to build in specific sensors and effectors in between the megamodels.

The paper is structured as follows:
Section~\ref{sec:megamodels:single-loop} discusses how megamodels are used to specify single feedback loops. 
Composing multiple feedback loops and especially their hierarchical composition are presented in Section~\ref{sec:megamodels:multiple-loops} and~\ref{sec:megamodels:multiple-loops:hierarchy}, respectively. 
Section~\ref{sec:metamodel+semantics} discusses the metamodel and the execution semantics for megamodels. 
The presented approach is discussed in Section~\ref{sec:evaluation+discussion}, while related work is reviewed in Section~\ref{sec:related-work}.
The paper concludes with Section~\ref{sec:conclusion+future-work}.

\section{Megamodels for Single Feedback Loops}
\label{sec:megamodels:single-loop}

This section discusses by illustrative examples how megamodels are used to specify single feedback loops and how a megamodel for a feedback loop can be modularized and composed. 
Megamodels specify a feedback loop by means of \emph{models}, \emph{model operations}, and the \emph{control flow} between the operations.
The modeling language shares characteristics with flowcharts and data flow diagrams: models are the data that represent, for example, the adaptable sub-system, and model operations that are organized in a control flow are computations that use and work on models. As an example, an operation can be an engine that checks invariants specified in a model on an architectural model of the system.

\subsection{Modeling a Single Feedback Loop}

\begin{figure}[!ht]
\centering
\includegraphics[width=1\columnwidth]{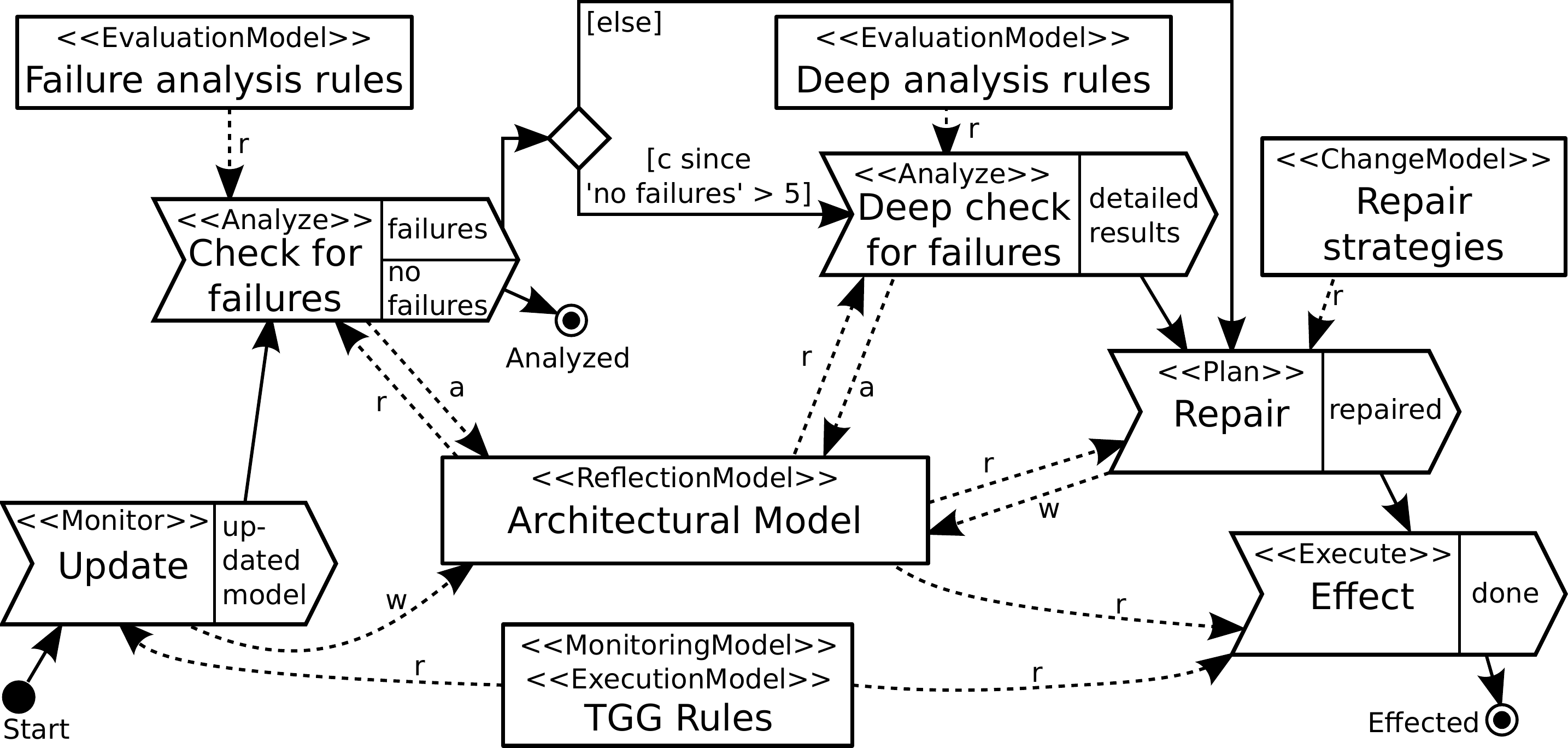}
\caption{Self-repair feedback loop}
\label{fig:self-repair}
\vspace{-1mm}
\end{figure}

Figure~\ref{fig:self-repair} depicts a megamodel that specifies a feedback loop for a self-repair scenario.
The concrete syntax for megamodels is as follows: a hexagon block arrow represents a \emph{model operation} that has one exit compartment for each return status of the operation. The rectangles depict \emph{models}, the solid arrows specify the \emph{control flow} between model operations that can be branched based on conditions, and the dotted arrows specify the use of models as \emph{inputs} or \emph{outputs} for model operations. To substantiate these elements for feedback loop concepts, they are labeled or stereotyped, which supports the modeler's perception of megamodels.

Model operations are assigned to the typical adaptation steps of a feedback loop: monitor, analyze, plan, and execute~\cite{Kephart&Chess2003}. Models are stereotyped based on the purpose they serve in self-adaptive software, which resulted from a categorization of runtime models we proposed in~\cite{VogelSG11}. \emph{Reflection models} represent the running software system to be adapted and the environment. These models are causally connected to the running system, i.e., relevant changes of the system are reflected in the model, and changes of the model are reflected in the system. Such models often represent the running system at the abstraction level of software architectures (cf.~\cite{VG10,georgas-computer09,GarCHSS04}). 
\emph{Evaluation models} and \emph{change models} describe how the running system is analyzed and how the system can be changed, respectively. Examples for such type of models are graph transformation rules in the form of \emph{Story Diagrams} or event-condition-action rules~\cite{VG12}.
Likewise, \emph{monitoring models} and \emph{execution models} support the monitor and execute steps, respectively. Finally, the use of models by model operations is substantiated to \emph{r}eading, \emph{w}riting, and \emph{a}nnotating models. While reading a model does not have any side effects, writes modify the model in a way that potentially affects the running system, and annotations to a model enrich a model without affecting the system.

With the megamodel depicted in Figure~\ref{fig:self-repair} we modeled an extended version of the self-repair scenario used in~\cite{VG10}. The \elem{Update} and \elem{Effect} operations use triple graph grammar rules (\elem{TGG Rules}) that specify by means of model transformation rules how the \elem{Architectural Model} is synchronized with the running system. Thus, based on observations of the running system, the \elem{Update} operation keeps the \elem{Architectural Model} up-to-date. The following analysis is conducted by the \elem{Check for failures} operation that employs \elem{Failure analysis rules} on the \elem{Architectural Model}. These rules define checks and constraints to identify failures. If no failures are identified, the feedback loop terminates. Otherwise, adaptation is required to repair these failures. 
At first, a decision is made whether further analysis is required. This is the case when the condition holds, which checks whether the last execution of the \elem{Check for failures} operation that has identified \elem{no failures} happened more than five consecutive executions in the past. Thus, the past five runs of the feedback loop were not able to repair the failures and a more thorough analysis may provide useful information for the planning step. This planning step uses the analysis results annotated by the previous operations to the \elem{Architectural Model} to select suitable \elem{Repair strategies}. The selected strategies change the \elem{Architectural Model} to prescribe a reconfiguration of the running system. This reconfiguration is executed to the running system by the \elem{Effect} operation that synchronizes the changes on the \elem{Architectural Model} to the running system and that finally terminates one run of the feedback loop.

This example illustrates how adaptation steps can be considered as abstract model operations that work on runtime models. Besides the control flow between multiple operations, the interplay between operations and runtime models has to be made explicit because the models are the basis for coordinating different model operations. Thus, this interplay is similar to dependencies between model operations, which are relevant for properly specifying and executing feedback loops with the help of megamodels.

\subsection{Modularizing and Composing Megamodels}

Specifying a more complex feedback loop also makes the related megamodel complex and hard to comprehend. To ease the modeling and perception of feedback loops based on megamodels, parts of a feedback loop can be abstracted, modeled in dedicated megamodels, and referenced by the megamodel specifying the whole feedback loop. Additionally, this supports the reusability of the abstracted feedback loop parts as they are specified in dedicated megamodels. Thus, besides specifying a complete feedback loop, a megamodel may also specify fragments of a feedback loop.

\begin{figure}[!ht]
\centering
\includegraphics[width=.75\columnwidth]{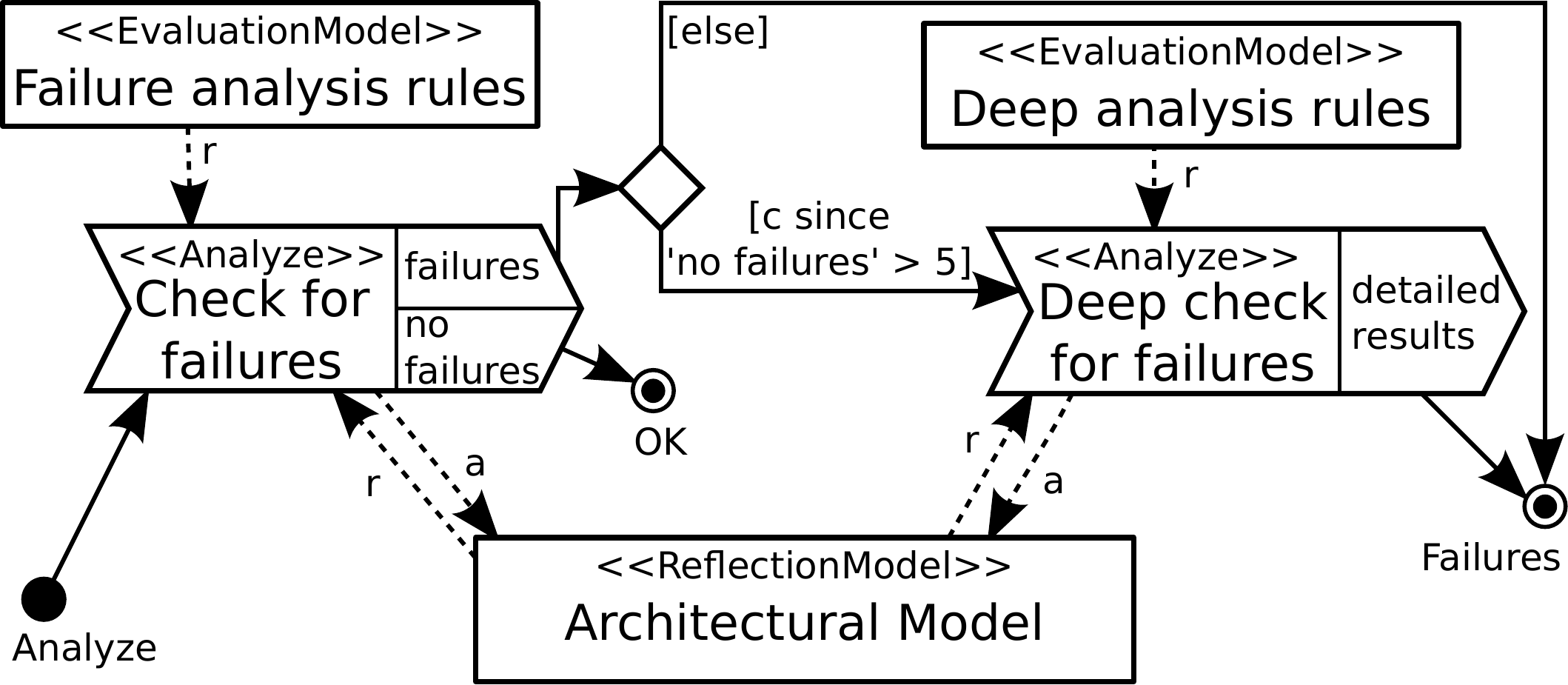}
\caption{Analysis step of the self-repair feedback loop}
\label{fig:self-repair-A}
\vspace{-1mm}
\end{figure}

For instance, the analysis step of the feedback loop for the self-repair example depicted in Figure~\ref{fig:self-repair} can be abstracted and specified in its own megamodel as shown in Figure~\ref{fig:self-repair-A}.

\begin{figure}[!ht]
\vspace{2mm}
\centering
\includegraphics[width=.85\columnwidth]{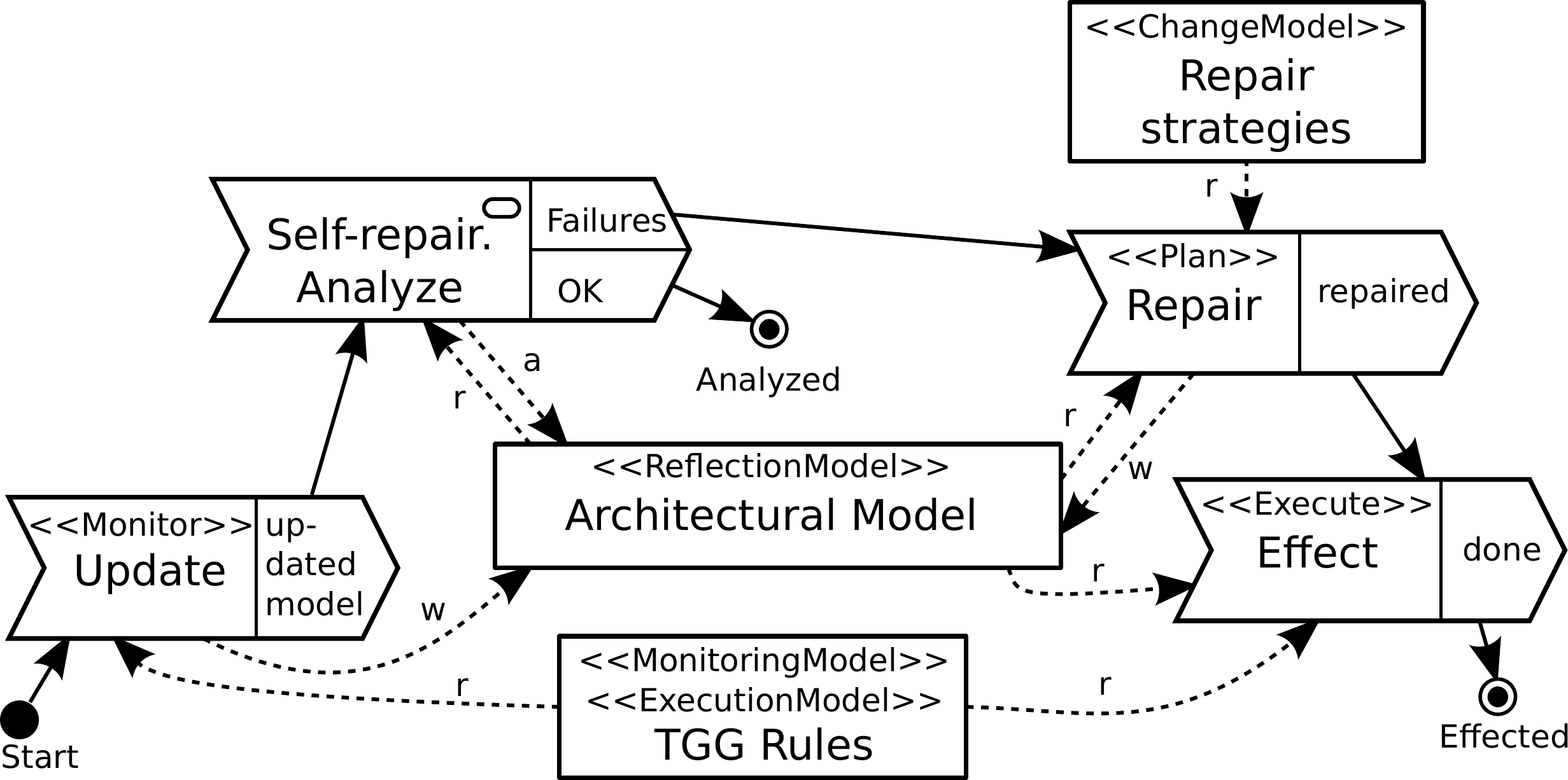}
\caption{Self-repair feedback loop using a complex model operation to invoke the analysis step defined in the megamodel shown in Figure~\ref{fig:self-repair-A}.}
\label{fig:self-repair-withoutA}
\vspace{-4mm}
\end{figure}

This analysis step of the self-repair feedback loop has one initial state called \elem{Analyze} and two final states reflecting whether failures have been identified (\elem{Failures}) or not (\elem{OK}). This megamodel can be (re)used and referenced by a different megamodel. Figure~\ref{fig:self-repair-withoutA} depicts the megamodel for the self-repair feedback loop that uses a \emph{complex model operation} to invoke the megamodel specifying the analysis step. The complex model operation is labeled with an icon, a small rounded rectangle, to distinguish it from the other type of model operations and to reveal that it references another megamodel. Based on the initial and final states of the invoked megamodel (cf.~Figure~\ref{fig:self-repair-A}), the complex model operation is named \elem{Self-repair.Analyze} and it provides two exit compartments, \elem{Failures} and \elem{OK}, respectively.
Thus, a complex model operation used in a megamodel abstracts from another megamodel and it synchronously invokes the abstracted megamodel when being executed.

Altogether, the specification of the self-repair feedback loop as shown in Figures~\ref{fig:self-repair-A} and~\ref{fig:self-repair-withoutA} is equivalent to the one shown in Figure~\ref{fig:self-repair}. The only difference is the number of megamodels used for the specification, which rather refers to technical modeling decisions than to design decisions. 

Providing the concept of complex model operations, the megamodel language supports the modular specification of a feedback loop.
For example, besides the analysis step all adaptation steps of a feedback loop can be specified in distinct megamodels, and a high-level megamodel comprising four complex operations for each step (monitor, analyze, plan, and execute) integrates all the corresponding megamodels.
Moreover, the depth of the abstraction and the related invocation relationships are not restricted.
In general, this leverages different abstraction levels for modeling feedback loops and it assists software engineers in modeling and understanding feedback loops. 

\section{Megamodels for Multiple Feedback Loops}
\label{sec:megamodels:multiple-loops}

In this section, we discuss the specification of multiple, interacting feedback loops by means of composing multiple megamodels each representing a feedback loop. Multiple feedback loops are required in a self-adaptive system if multiple concerns, like failures or performance, have to be addressed. Each concern requires its specific models and model operations. Therefore, different self-management capabilities, like self-healing or self-optimization, are realized by different feedback loops. However, the feedback loops have to coordinate their adaptations because of competing concerns. Otherwise, adaptations of different feedback loops might conflict each other. For example, an adaptation of the system to optimize performance might cause failures in the managed system, or healing a failure might degrade performance.
Thus, coordination between multiple feedback loops is required, which is done by integrating or composing multiple megamodels each specifying a feedback loop.

\begin{figure}[!ht]
\vspace{2mm}
\centering
\includegraphics[width=.85\columnwidth]{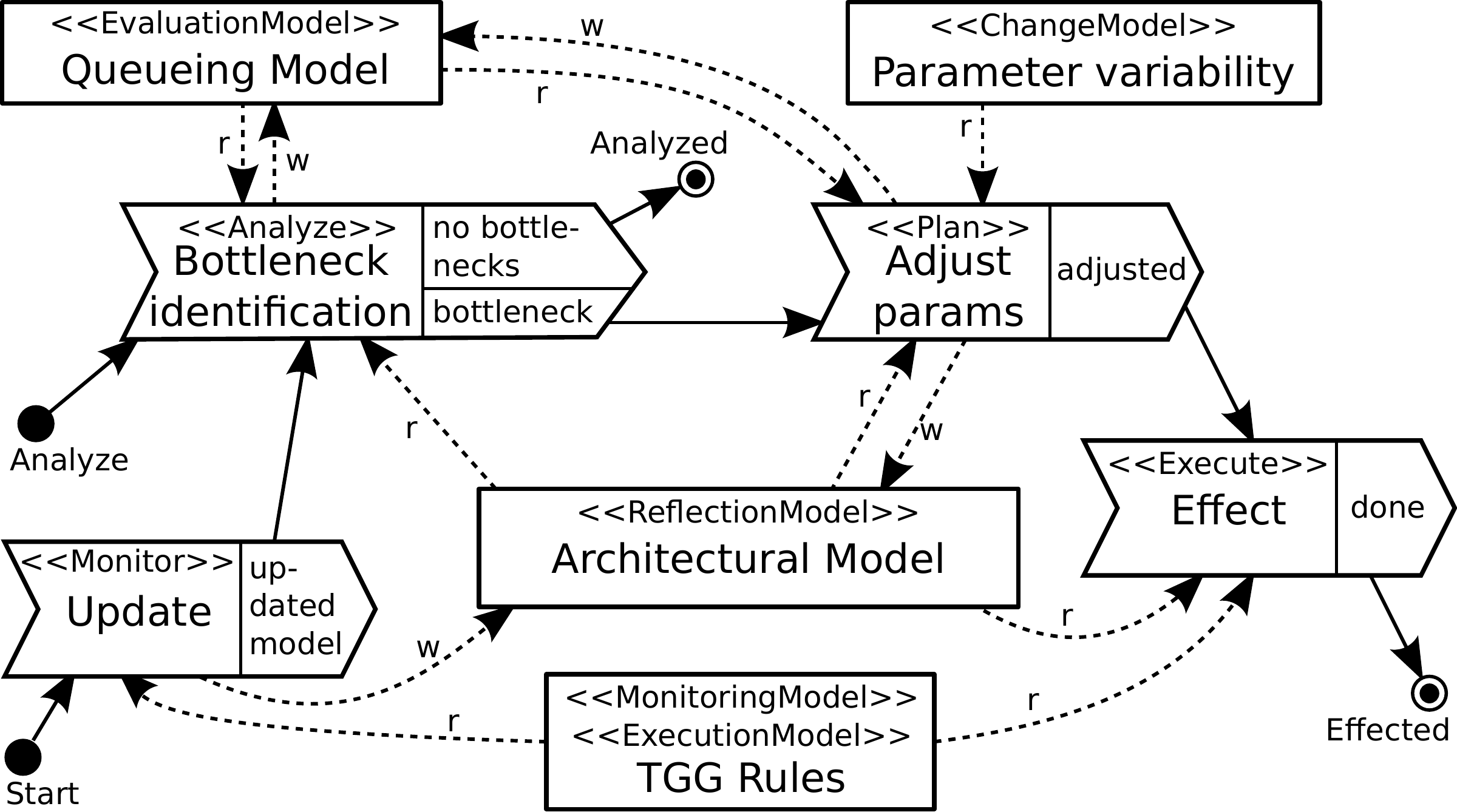}
\vspace{-1mm}
\caption{Self-optimization feedback loop}
\label{fig:self-optimization}
\vspace{-6mm}
\end{figure}

In the following, we discuss the composition of two feedback loops using the megamodel language. Besides the self-repair feedback loop depicted in Figure~\ref{fig:self-repair} or in Figures~\ref{fig:self-repair-A} and~\ref{fig:self-repair-withoutA}, the self-adaptive system should be equipped with a self-optimization feedback loop as shown in Figure~\ref{fig:self-optimization}.

From the modeling perspective, the self-optimization feedback loop is quite similar to the self-repair loop. It has to be noted that both feedback loops work on the same instance of the \elem{Architectural Model}, and thus the monitoring and the execution steps of both loops provide the same functionality of synchronizing the \elem{Architectural Model} with the running system.
A difference to the self-repair feedback loop is that the self-optimization feedback loop has two initial states either initiating the loop with the monitoring step or with the analysis step. Moreover, in contrast to the self-repair feedback loop, the analysis and planning steps of the self-optimization feedback loop do not only use the \elem{Architectural Model} to exchange information but additionally the \elem{Queueing Model}, either to identify bottlenecks in the running system or to identify reasonable values for parameters given by the \elem{Parameter variability} model to adjust the configuration of the managed system in order to resolve bottlenecks.

Using two basic coordination mechanisms that linearize either complete feedback loops or just the analysis and planning steps of different feedback loops, we discuss below how the megamodel language is applied to model the composition of multiple, interacting feedback loops.

\subsection{Linearizing Complete Feedback Loops}

\begin{figure}[!ht]
\vspace{-1mm}
\centering
\includegraphics[width=1\columnwidth]{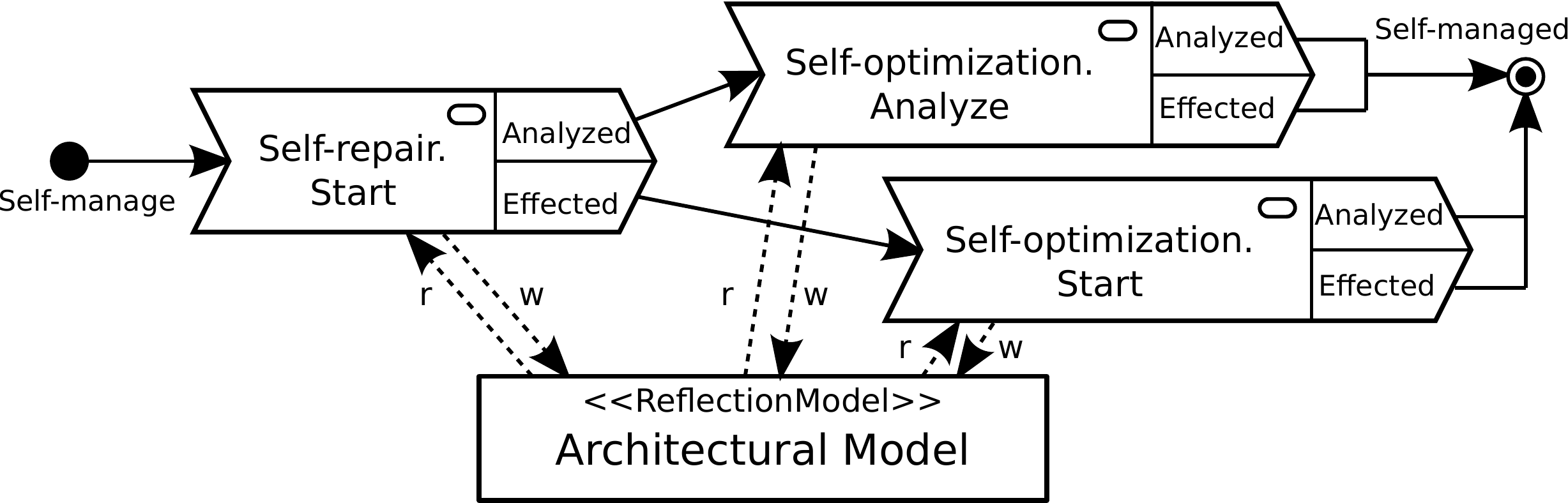}
\vspace{-6mm}
\caption{Linearizing complete feedback loops by invoking the self-repair loop (cf.~Figure~\ref{fig:self-repair-withoutA}) followed by the self-optimization loop (cf.~Figure~\ref{fig:self-optimization}).}
\label{fig:loop-sequence}
\vspace{-3mm}
\end{figure}

A simple way to coordinate two feedback loops is to completely linearize them. This is described in the megamodel depicted in Figure~\ref{fig:loop-sequence}, which uses complex model operations to synchronously invoke the individual loops. Both feedback loops share the same instances of the \elem{Architectural Model} and \elem{TGG Rules}. To keep the megamodel illustrative, the \elem{TGG Rules} are not depicted in Figure~\ref{fig:loop-sequence} since they, in contrast to the architectural model, do not change at runtime.

In this example, a higher priority is assigned to repairing failures than to optimizing the performance because failures are often more harmful than slow response times. Moreover, optimizing the performance of a failing system before the failures have been repaired is not reasonable.
Therefore, the self-repair feedback loop is executed before the self-optimization loop. In Figure~\ref{fig:loop-sequence}, \elem{Self-repair.Start}  invokes the self-repair feedback loop as specified by the megamodel shown in Figure~\ref{fig:self-repair-withoutA}. 
Thus, the monitoring and analysis steps are carried out, while the first one updates the \elem{Architectural Model} to reflect the current state of the running system, and the latter one analyzes this model for failures. Depending on whether failures have been found or not, the feedback loop either continues  with the following adaptation steps or it terminates, respectively. This influences the subsequent execution of the self-optimization feedback loop.

If no failures have been identified, the self-repair feedback loop does not need to plan and execute any adaptations and it terminates in the state \elem{Analyzed}. The subsequent self-optimization feedback loop may immediately start with the analysis step because the monitoring step of the previous self-repair loop already updated the shared \elem{Architectural Model} and no adaptations have been performed by the self-repair loop. Thus, the complex model operation \elem{Self-optimization.Analyze} in Figure~\ref{fig:loop-sequence} invokes the self-optimization loop that begins execution in the initial state \elem{Analyze} (cf.~Figure~\ref{fig:self-optimization}). If no bottlenecks have been identified, the self-optimization feedback loop terminates in the state \elem{Analyzed}. Otherwise, it carries out the planning and execution steps, and terminates in the state \elem{Effected}.

On the other hand, if the self-repair feedback loop has identified failures, it plans and executes changes to the running system and it terminates in the state \elem{Effected} (cf.~Figure~\ref{fig:self-repair-withoutA}). Thus, the running system has been modified, which requires that the subsequent self-optimization feedback loop performs the monitoring step. Thus, the self-optimization loop is invoked by the complex model operation \elem{Self-optimization.Start} shown in Figure~\ref{fig:loop-sequence} to begin execution in the initial state \elem{Start} (cf. Figure~\ref{fig:self-optimization}). After carrying out the monitoring and analyzing steps, the self-optimization feedback loop either terminates or, if required, performs the planning and execution steps similar to the previous case.

This coordination mechanism synchronizes different feedback loops by sequentially executing them and by using the running system if one loop performs adaptations. Thus, adaptations performed by a feedback loop are executed to the system before another loop performs its adaptation steps. If a feedback loop does not perform any adaptations, the subsequent loop may start right away with the analysis step.

\subsection{Linearizing Analysis and Planning of Feedback Loops}

\begin{figure}[!ht]
\vspace{2mm}
\centering
\subfloat[Self-repair]{\includegraphics[width=.9\columnwidth]{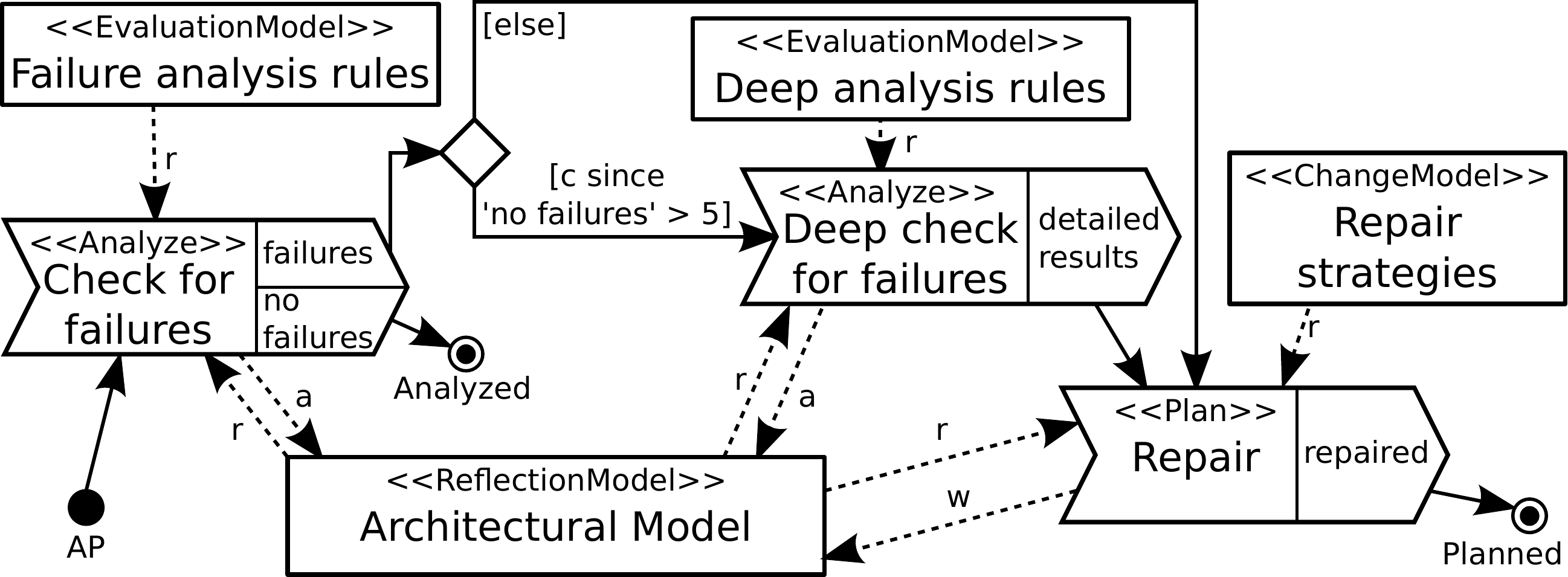}
\label{fig:self-repair-AP}}
\vfil
\subfloat[Self-optimization]{\includegraphics[width=.8\columnwidth]{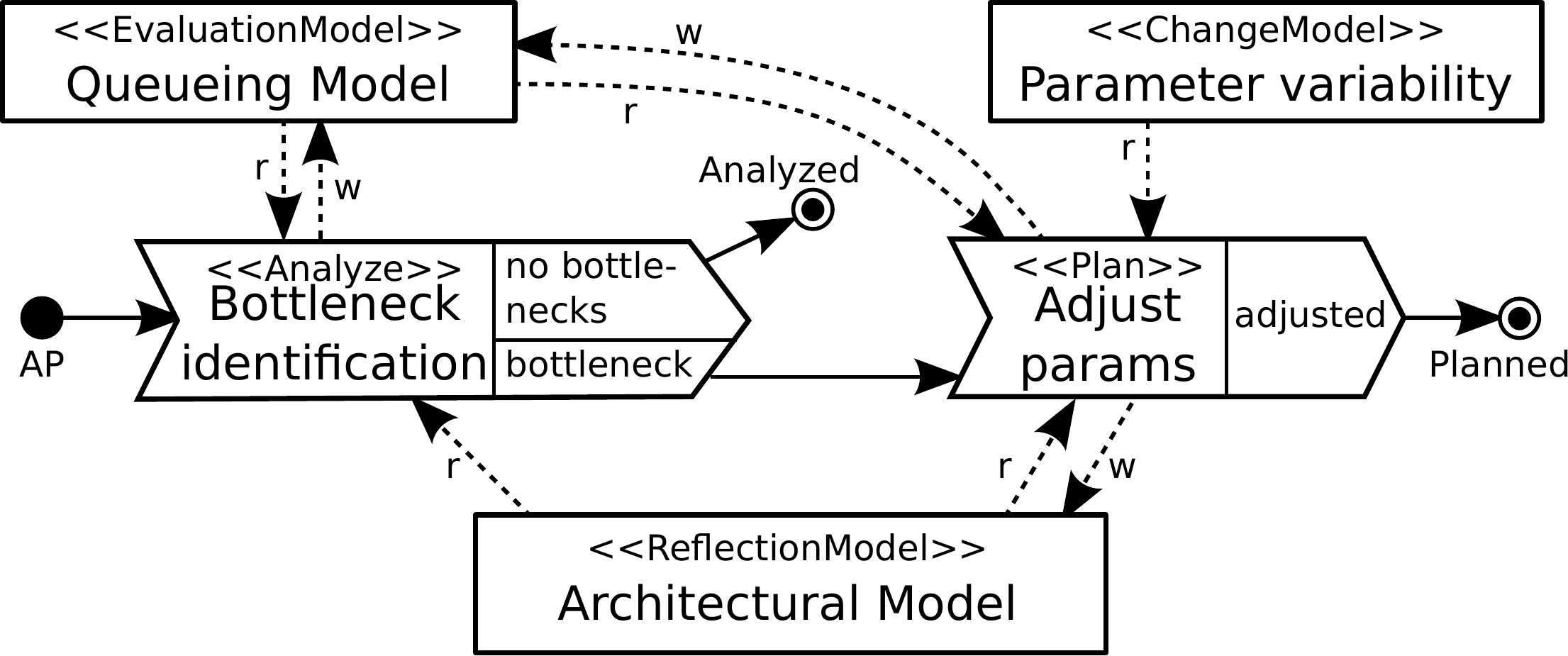}
\label{fig:self-optimization-AP}}
\caption{Analysis and planning for \protect\subref{fig:self-repair-AP} self-repair, \protect\subref{fig:self-optimization-AP} self-optimization}
\label{fig:AP}
\vspace{-4mm}
\end{figure}

The other way to coordinate multiple feedback loops, which we want to present, is to synchronize the feedback loops in common monitoring and execution steps, while the analysis and planning steps are linearized. 
Therefore, in the example of coordinating the feedback loops for self-repairing (cf.~Figures~\ref{fig:self-repair} or~\ref{fig:self-repair-withoutA}) and self-optimization (cf.~Figure~\ref{fig:self-optimization}), the analysis and planning steps of each of these loops are specified in dedicated megamodels as shown in Figure~\ref{fig:AP}. 

In simple terms, the analysis and planning steps have just been cut out from the above megamodels and no changes have been done to their specifications.
Alternatively, the above megamodels could have been extended with appropriate initial and final states to enter a megamodel at the analysis step and to exit a megamodel after the planning step. However, to keep the megamodels simple, we specified the analysis and planning steps in dedicated megamodels.

\begin{figure}[!ht]
\centering
\includegraphics[width=.9\columnwidth]{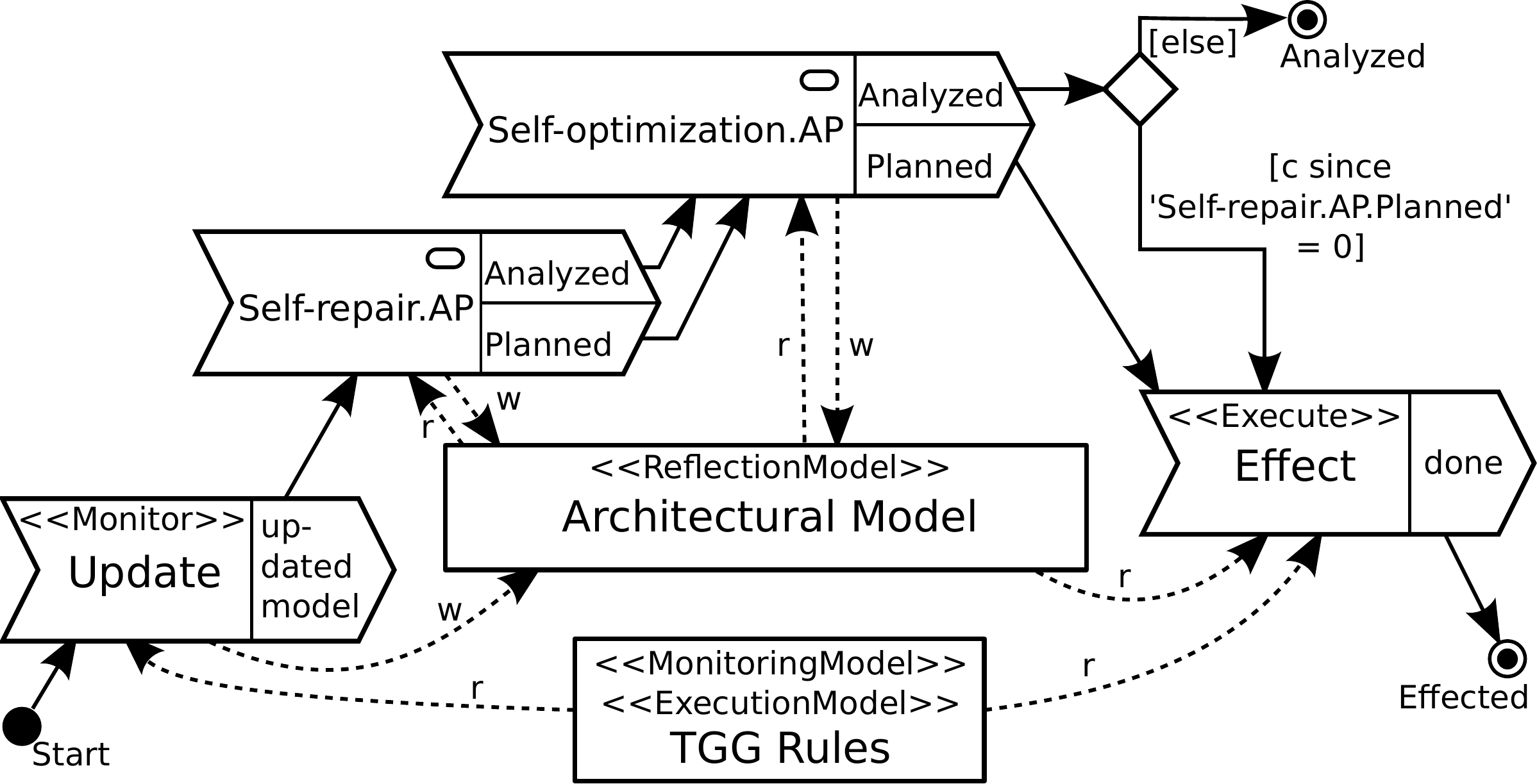}
\caption{Linearizing the analysis and planning steps of the self-repair (cf.~Figure~\ref{fig:self-repair-AP}) and self-optimization (cf.~Figure~\ref{fig:self-optimization-AP}) feedback loops.}
\label{fig:loop-overlapping}
\vspace{-1mm}
\end{figure}

The coordination of the self-repair and self-optimization feedback loops, specifically of their analysis and planning steps, is specified in the megamodel depicted in Figure~\ref{fig:loop-overlapping}.

Likewise to the previous example, the \elem{Update} and \elem{Effect} operations synchronize the \elem{Architectural Model} with the running system respectively for monitoring and for executing changes. Since the self-repair and the self-optimization feedback loops work on the same instance of the \elem{Architectural Model}, they share the monitoring and execution steps.
However, the analysis and planning steps are specific for the addressed concerns, like failures or performance, and they decide if and how the system should be adapted. Thus, they must coordinate each other to tackle competing concerns and as a consequence potentially conflicting adaptations.

Therefore, the analysis and planning steps for self-repairing failures are executed before the analysis and planning steps for self-optimizing performance. This is modeled in Figure~\ref{fig:loop-overlapping} by complex model operations sequentially and synchronously invoking the megamodels specifying the analysis and planning steps for both concerns as depicted in Figure~\ref{fig:AP}.
The \elem{Architectural Model} is only modified by the self-repair's planning step if the related analysis step has identified failures in the running system. These modifications are planned adaptations in order to repair the failures, and they have been applied at the model level but they have not been effected to the running system.

The subsequent analysis and planning steps of the self-optimization feedback loop use the \elem{Architectural Model} that has been potentially modified by the self-repair's planning step. If the model has not been modified, there are no conflicting adaptations. Otherwise, the adaptations proposed by the self-repair feedback loop must be handled by the self-optimization loop. Two scenarios are possible here. 

First, the proposed adaptations are considered as invariants that the self-optimization's analysis and planning steps must adhere to. Thus, a proposed adaptation by the self-optimization loop is not allowed to change the system in a way that contradicts the proposed adaptation by the preceding self-repair loop. In this case, the healing of failures is assigned a higher priority than optimizing performance.

Second, the self-optimization's analysis and planning steps may just override the adaptations proposed by the preceding self-repair feedback loop. In this case, the priorities are inverted since the self-optimization's analysis and planning steps do not have to incorporate the adaptation proposed by the self-repair feedback loop.

Considering Figure~\ref{fig:loop-overlapping}, when the self-optimization's analysis and planning steps terminate, the \elem{Effect} operation is executed if adaptations are proposed in the \elem{Architectural Model} by the self-repair's or the self-optimization's planning steps. Thus, at least one of the complex model operations \elem{Self-repair.AP} or \elem{Self-optimization.AP} must terminate in the state \elem{Planned}. Otherwise, the megamodel depicted in Figure~\ref{fig:loop-overlapping} terminates in the state \elem{Analyzed} since no failures and no bottlenecks have been identified, which does not require any adaptation to be planned and executed to the running system.

As depicted in the Figures~\ref{fig:loop-sequence} and~\ref{fig:loop-overlapping}, both megamodel examples and their usage of complex model operations show how the coordination of multiple feedback loops can be specified at a high-level of abstraction. The coordination is achieved by executing interacting adaptation steps of different loops in a controlled manner by sequential and synchronous invocations that are explicitly specified in the megamodels.
Besides megamodels specifying single feedback loops or individual adaptation steps of feedback loops (cf.~Figures~\ref{fig:self-repair},\,\ref{fig:self-repair-A},\,\ref{fig:self-repair-withoutA},\,\ref{fig:self-optimization}, or~\ref{fig:AP}), distinct megamodels as shown in Figures~\ref{fig:loop-sequence} and~\ref{fig:loop-overlapping} may explicitly specify the coordination of multiple, interacting feedback loops. Thus, the same modeling language is used to describe single feedback loops as well as the coordination of multiple feedback loops.

\section{Hierarchy of Feedback Loops}
\label{sec:megamodels:multiple-loops:hierarchy}

In this section, we describe the particular case of composing multiple feedback loops in hierarchies. This is required for adaptive control architectures, like the layered architecture for self-managed system proposed in~\cite{Kramer&Magee2007} or hierarchical structures with internal layers as presented in~\cite{Hestermeyer+2004}.

The basic idea is that a running system is managed by multiple feedback loops that are organized in layers. Even if only one specific self-management capability, like self-healing, is supported by a self-adaptive system, multiple feedback loops can be employed simultaneously. Different time scales of feedback loops are a basic criteria for placing feedback loops in different layers (cf.~\cite{Kramer&Magee2007}). A feedback loop realizing urgent or frequent adaptations must work at shorter time scales and thus, it is placed at a lower layer. In contrast, a feedback loop performing long-term or complex planning that is rather rarely required often work at longer time scales, which places the loop at a higher level.

A particular aspect of layering feedback loops is that a feedback loop at a certain layer is managed itself by the feedback loop at the adjacent, higher level layer. Considering the running system being located at $layer_0$, the system is directly managed by the feedback loop located at $layer_1$. A feedback loop at $layer_2$ directly manages the feedback loop at $layer_1$ and thus, indirectly the running system at $layer_0$. While three layers are proposed in~\cite{Kramer&Magee2007}, there is theoretically no limit for the number of layers. 
However, managing a feedback loop by another feedback loop requires integrating sensors and effectors for the managed loop. This enables the managing loop to reflect the managed loop in order to perform its adaptation steps. This is similar to the \emph{reflection models} used in the previous examples that reflect the running system and that serve as a basis for the adaptation steps. 

In the following, we show how a megamodel specifying a managed feedback loop directly serves as a reflection model for the managing loop. Thus, this megamodel is an abstraction of the managed loop for the managing loop, and this megamodel is just a model in the managing loop's megamodel. 
This is possible because megamodels are kept alive at runtime and they specify a feedback loop by means of model operations, the control flow between operations, and the models used by operations. Hence, a managing feedback loop may use a megamodel reflecting a managed loop in order to adapt the model operations, control flow, and models of the managed loop. Following an interpreter approach to execute megamodels, the flexibility required for coping with megamodel changes at runtime is provided. 

\begin{figure}[!t]
\vspace{2mm}
\centering
\includegraphics[width=1\columnwidth]{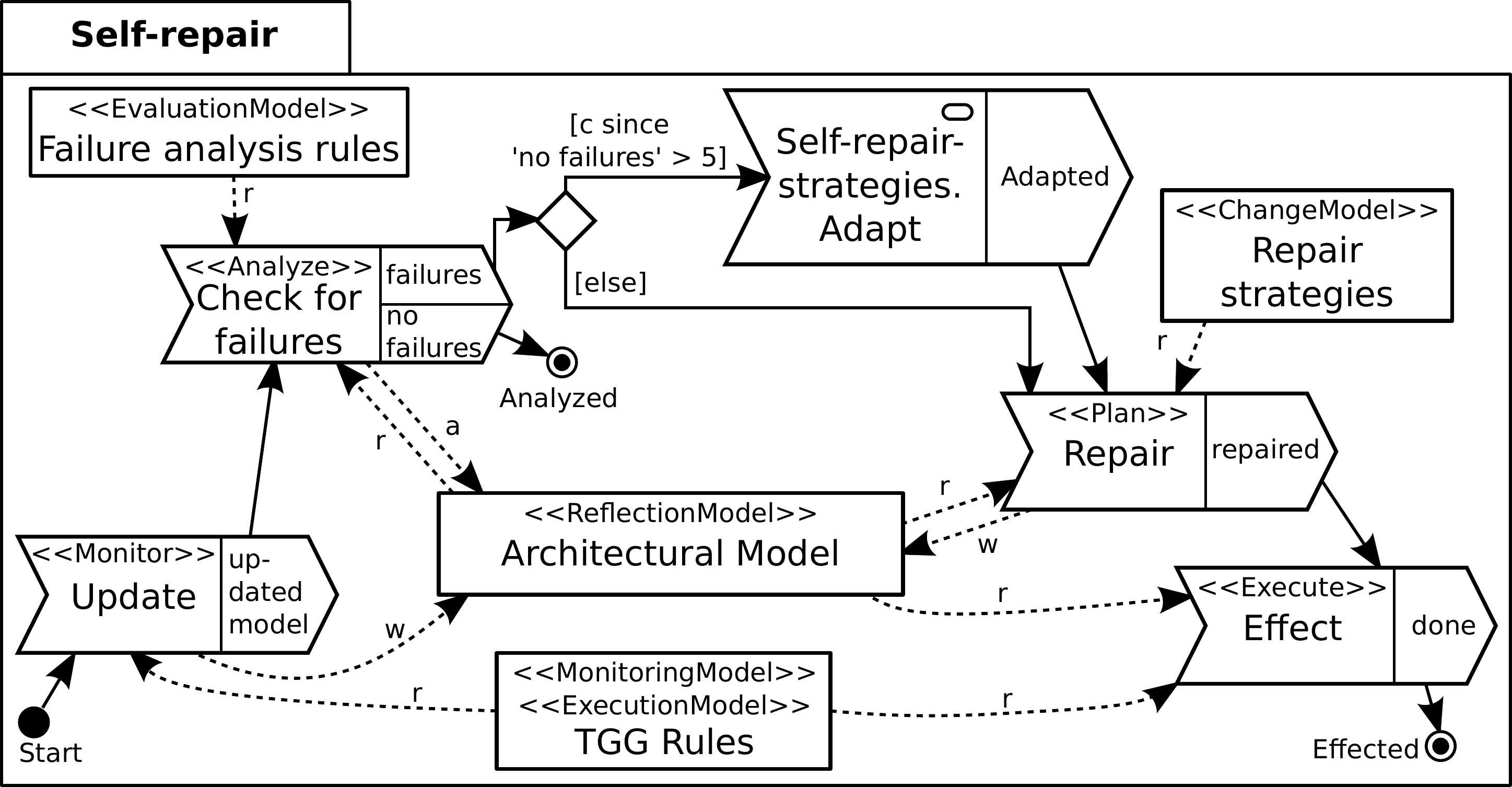}
\caption{\elem{Self-repair} at $layer_1$}
\label{fig:self-repair-layer1}
\vspace{-5mm}
\end{figure}

\begin{figure}[!ht]
\vspace{-1mm}
\centering
\includegraphics[width=.9\columnwidth]{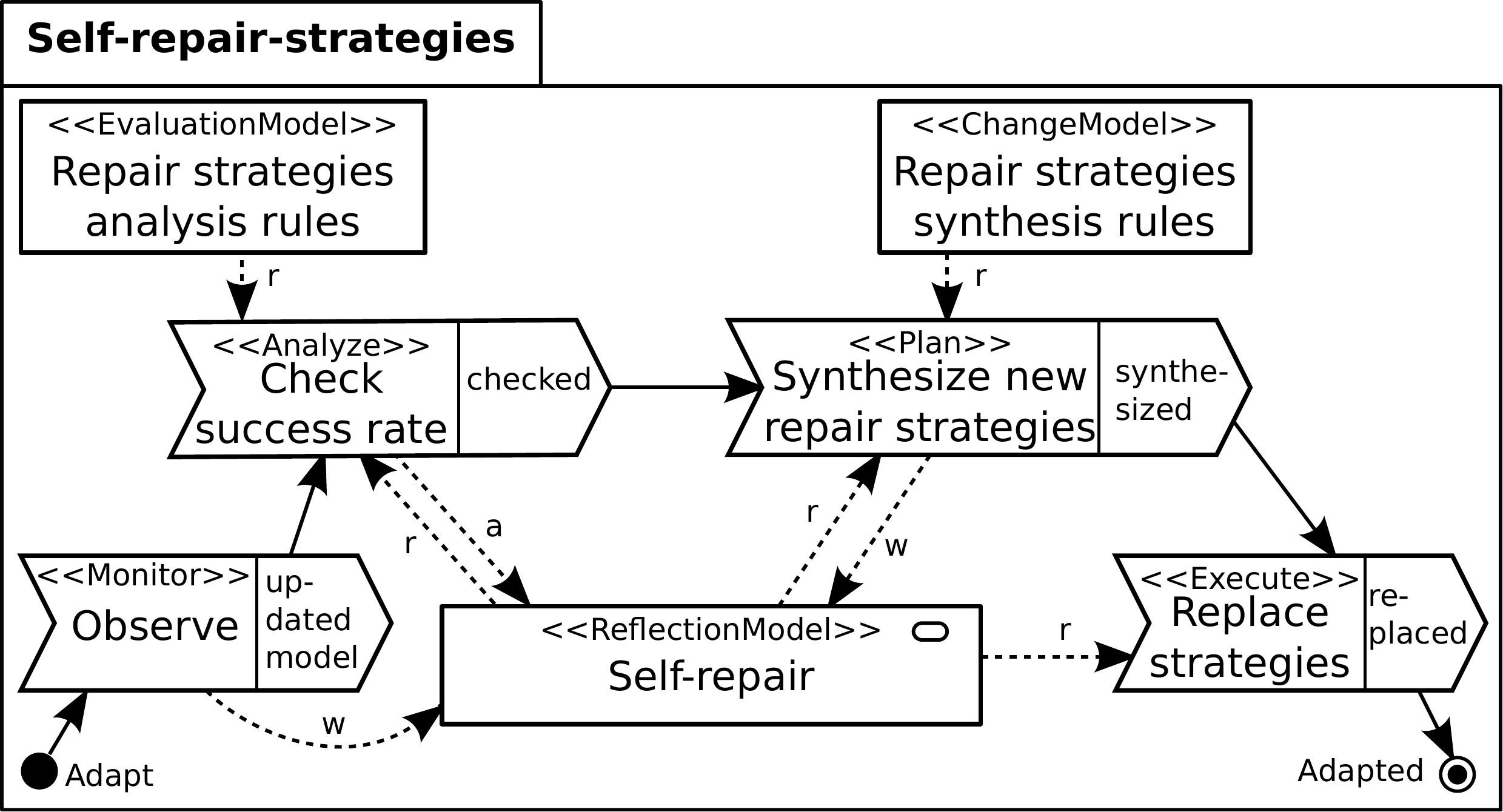}
\caption{\elem{Self-repair-strategies} at $layer_2$}
\label{fig:self-repair-layer2}
\vspace{-1mm}
\end{figure}

This is illustrated by the following example. Figure~\ref{fig:self-repair-layer1} shows the megamodel specifying a self-repair feedback loop similar to the one shown in Figure~\ref{fig:self-repair-withoutA}. This megamodel is named \elem{Self-repair} and it is located at $layer_1$. Thus, this feedback loop directly manages the running system that is reflected in the \elem{Architectural Model} by checking for failures and repairing them using pre-defined \elem{Repair strategies}.\, 
However, these repair strategies need not to be able to handle all kinds of failures. This would require that all kinds of failures could have been anticipated when developing and deploying these strategies, which is usually not the case given the uncertainty concerning self-adaptive systems and their environments. Thus, the repair strategies have to be maintained and adapted at runtime, and this task can be assigned to another feedback loop located at $layer_2$. This feedback loop is specified by the megamodel \elem{Self-repair-strategies} depicted in Figure~\ref{fig:self-repair-layer2}.

This megamodel is triggered by the $layer_1$ feedback loop depicted in Figure~\ref{fig:self-repair-layer1} using the complex operation \elem{Self-repair-strategies.Adapt} when more than five of the last consecutive runs of the $layer_1$ loop were not able to repair the failure. Thus, the last execution of the \elem{Check for failures} operation that has identified \elem{no failures} happened more than five consecutive executions in the past (cf. condition in Figure~\ref{fig:self-repair-layer1}). This indicates that the current repair strategies are not able to heal the failure and new or adjusted strategies are required.

Therefore, the \elem{Self-repair-strategies} feedback loop comes into play at $layer_2$ and manages the \elem{Self-repair} loop located at $layer_1$.
The megamodel itself that defines the \elem{Self-repair} loop is directly used as the reflection model to perform the \elem{Self-repair-strategies}' adaptation steps (cf. \elem{$\ll$ReflectionModel$\gg$} in~Figure~\ref{fig:self-repair-layer2} labeled with an icon, a small rounded rectangle, to highlight that this model is a megamodel).
As a consequence, the \elem{Self-repair-strategies} feedback loop directly works on the \elem{Self-repair} megamodel to check the success rates of the existing strategies, to synthesize new strategies, and finally, to replace the existing strategies with the new ones. The last step adapts the \elem{Self-repair} loop by linking a new \elem{Repair strategies} model into the megamodel specifying the \elem{Self-repair} loop. This adaptation equips the \elem{Self-repair} feedback loop with new repair strategies to be used from now on.

Another adaptation the $layer_2$ feedback loop may perform is to modify the control flow of the $layer_1$ loop. This is not specified in the megamodel shown in Figure~\ref{fig:self-repair-layer2}, but for example, the \elem{Self-repair-strategies} loop may increase or decrease the constant $5$ in the condition used in the \elem{Self-repair} loop depicted in Figure~\ref{fig:self-repair-layer1}. This changes the \elem{Self-repair} loop's control flow at $layer_1$ and it influences how often the \elem{Self-repair-strategies} loop at $layer_2$ is triggered and executed.
In general, besides changing the expressions of conditions, the control flow can be adjusted by changing model operations and the way they are linked to each other. 

These examples illustrate that by using a megamodel specifying a feedback loop at $layer_n$ directly as the reflection model for the feedback loop at $layer_{n+1}$, hierarchies of feedback loops can be easily built without having to integrate specific sensors or effectors. In other terms, a megamodel specifying and executing a feedback loop is directly used to dynamically adapt this feedback loop. This is possible because megamodels are explicitly kept alive at runtime and they are executed by an interpreter that provides the required flexibility to cope with changes of a megamodel at runtime.

\section{Metamodel and Execution Semantics}
\label{sec:metamodel+semantics}

\begin{figure*}[!ht]
\vspace{2mm}
\centering
\includegraphics[width=.84\textwidth]{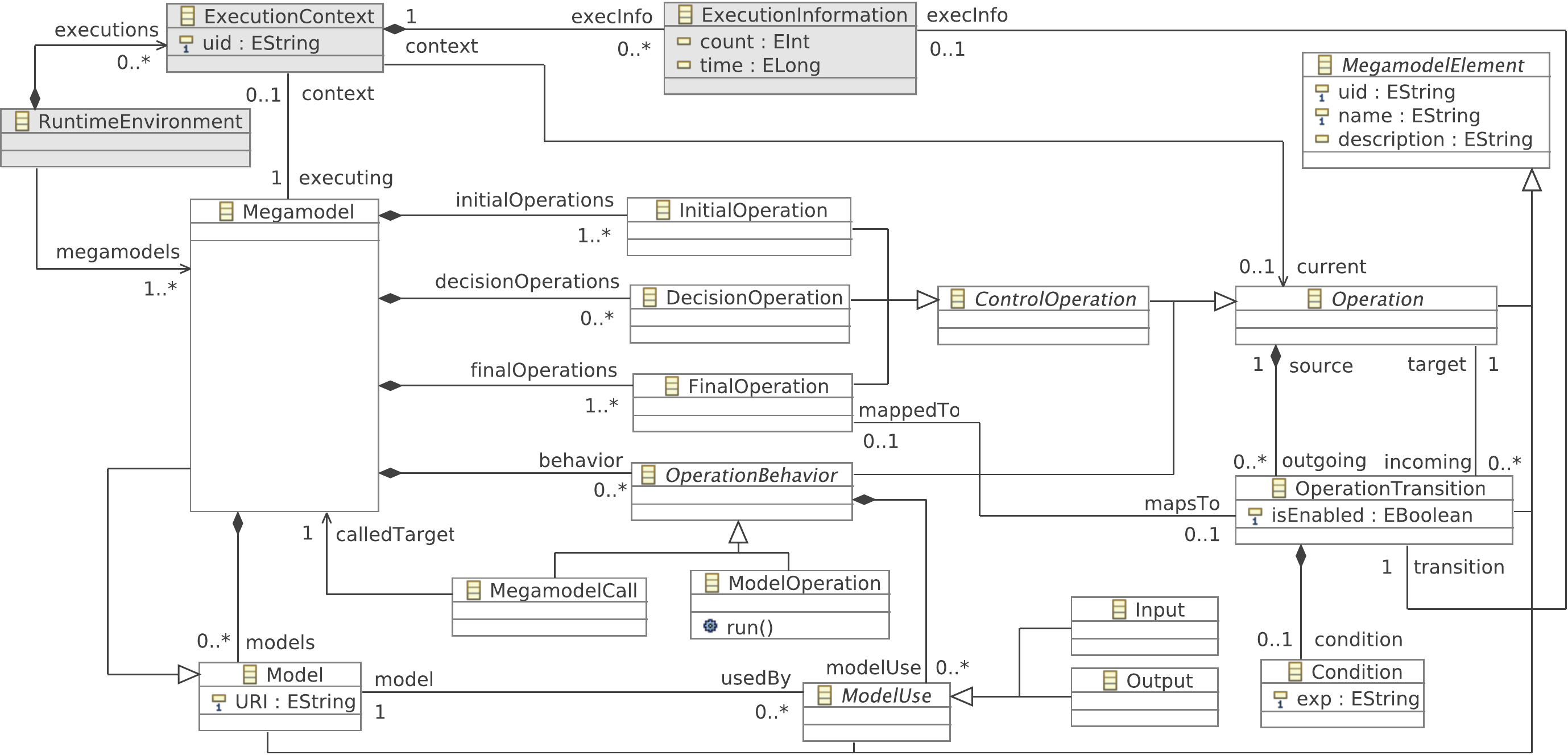}
\vspace{-1mm}
\caption{Metamodel for modeling and executing megamodels}
\label{fig:metamodel}
\vspace{-5mm}
\end{figure*}

In this section, we discuss the metamodel defining the modeling language by means of the abstract syntax for megamodels, as well as the execution semantics that are relevant for the interpreter to execute megamodels. Conceptually, a megamodel describes a feedback loop by means of operations, the control flow between operations, and the models that are used by operations.

The metamodel is depicted in Figure~\ref{fig:metamodel}, whose wide-shaded elements show concepts relevant for modeling megamodels and whose gray-shaded elements describe concepts relevant for executing megamodels.
The core concept is the \elem{Megamodel} that contains \elem{Model}s and different kinds of \elem{Operation}s. Operations are linked with each other by \elem{OperationTransition}s that define the control flow between operations. Each \elem{OperationTransition} connects exactly one \elem{source} to one \elem{target} operation.

The different \elem{Operation} types used in a \elem{Megamodel} are the following:
at least one \elem{InitialOperation} and at least one \elem{FinalOperation} are required, which are the megamodel's initial and final states, respectively. \elem{DecisionOperation}s allow us to branch the control flow based on \elem{Condition}s that are annotated to a \elem{DecisionOperation}'s \elem{outgoing OperationTransition}s (see Figures~\ref{fig:self-repair} or~\ref{fig:self-repair-layer1} for an example).
The megamodel's actual behavior is defined by \elem{OperationBehavior}s, either \elem{ModelOperation}s or \elem{MegamodelCall}s, that use {Model}s as \elem{Input} or \elem{Output}. \elem{ModelOperation}s are atomic computation units whose implementations are triggered by the \elem{run} method. \elem{MegamodelCall}s are the complex operations, as used in the examples of the previous sections, to invoke a \elem{Megamodel} that specifies either parts of the same feedback loop or a complete and different feedback loop. This enables modular megamodels for describing single feedback loops (cf. Section~\ref{sec:megamodels:single-loop}) and the coordination of multiple loops (cf.~Sections~\ref{sec:megamodels:multiple-loops}/\ref{sec:megamodels:multiple-loops:hierarchy}). Therefore, each \elem{outgoing OperationTransition} of a \elem{MegamodelCall} is \elem{mappedTo} to a \elem{FinalOperation} of the called \elem{Megamodel} to properly proceed the execution when the execution of the invoked megamodel has finished.

Since a \elem{Megamodel} is itself a \elem{Model}, it can be part of another megamodel and just be used similar to a ``normal'' model. For example, the megamodel specifying the \elem{Self-repair-strategies} feedback loop shown in Figure~\ref{fig:self-repair-layer2} contains the megamodel representing the \elem{Self-repair} feedback loop and this contained megamodel is used by the adaptation steps, i.e., the operations of the \elem{Self-repair-strategies} loop.

Finally, the abstract class \elem{MegamodelElement} in the metamodel (cf.~Figure~\ref{fig:metamodel}) provides identifiers, names, and descriptions for the main megamodel concepts.

Concerning the execution semantics that are relevant for the megamodel interpreter, the gray-shaded metamodel elements come into play. The \elem{RuntimeEnvironment} manages the execution of megamodels by maintaining an \elem{ExecutionContext} for each megamodel, which points to the \elem{current}ly executed \elem{Operation}. Moreover, the context maintains \elem{ExecutionInformation}, especially \elem{count} and \elem{time}, for each \elem{OperationTransition}, i.e., for each link connecting operations in the corresponding megamodel. While \elem{time} represents a timestamp when the transition has been taken the last time, \elem{count} reflects the number how often the source operation of the transition has been executed without taking the corresponding transition but another outgoing transition. If the corresponding transition is taken, \elem{count} is reset to $0$. This information is maintained by the interpreter and used in expressing \elem{Condition}s that are evaluated by the interpreter. For example, a counter can be compared with a constant to branch the control flow in a megamodel as it is shown in Figures~\ref{fig:self-repair} or~\ref{fig:self-repair-layer1}. For conditions, basic arithmetic and boolean operations on \elem{time} and \elem{count} are currently supported. 

This expression language for conditions is kept generic to clearly separate the abstraction levels between a megamodel and its contained models and operations. The interpreter works at the level of megamodels and it considers the individual models and operations as black boxes. Since conditions are evaluated by the interpreter, they must not utilize internal concepts of the models or operations. Otherwise, it would require that the interpreter must access such concepts, which would couple the interpreter implementation to the specific implementations of the models or operations. In the end, this would prevent reuse of the interpreter for different self-adaptive systems. However, if more advanced or application-specific conditions are needed, they can be modeled by appropriate \elem{outgoing OperationTransition}s that can be seen as return states of operations and for which \elem{time} and \elem{count} are maintained. In this case, the operation's implementation decide on the state the operation terminates in, while the interpreter may further branch the control flow based on the \elem{time} and \elem{count} attributes of the return state.

Concerning the execution of megamodels by the interpreter, a simple way has been chosen. Each megamodel is handled as a singleton that is reused if the megamodel is executed multiple times, e.g., it can be invoked multiple times from the same or from different locations in other megamodels. This avoids issues on creating and destroying megamodel instances.
Moreover, \elem{MegamodelCall}s and \elem{ModelOperation}s are executed synchronously and there is always only one active thread of control. Thus, concurrent executions of megamodels or operations are not supported and thus, no synchronization mechanisms are required.

The metamodel has been implemented with the \emph{Eclipse Modeling Framework}~(EMF) and the dynamic EMF capabilities avoid the need to generate code for the metamodel, which supports the megamodel interpreter.
The stereotypes, like \elem{$\ll$ReflectionModel$\gg$} or \elem{$\ll$Monitor$\gg$}, and labels, like \elem{r}, \elem{a}, or \elem{w}, used for \elem{Model}s, \elem{ModelOperation}s, or \elem{ModelUse}s, which have been introduced in Section~\ref{sec:megamodels:single-loop}, are not directly supported by the metamodel. However, they do not influence the execution semantics and thus, they are not relevant for the interpreter. However, they can be introduced in the language by extending the metamodel or by using a profile. 

\section{Evaluation and Discussion}
\label{sec:evaluation+discussion}

This section evaluates and discusses our domain-specific modeling solution for feedback loops in self-adaptive systems. This includes the language to specify feedback loops with megamodels and the interpreter to execute megamodels.

Feedback loops are crucial elements in the architecture of a self-adaptive system and they should be made explicit in the design and analysis of the self-adaptive system~\cite{Brun+2009,PezzeMS08}.
Using the modeling language we propose in this paper, a feedback loop is explicitly specified in the megamodels. Rather than treating the adaptation logic realizing a feedback loop as a black box component, our language considers individual adaptation steps, like monitoring, analysis, planning, and execution, and how these steps cooperate to form a feedback loop. Thereby, adaptation steps can be modeled in distinct megamodels and used in other megamodels that integrate the different steps. This supports the reusability of megamodels as well as the modular specification of feedback loops, which eases the development of the adaptation logic.

This modularity also leverages specifying the composition of feedback loops in a homogeneous manner. Megamodels in our language are not only used to specify single feedback loops but also the composition of multiple feedback loops.

Thereby, the modeling language targets a reasonable abstraction level similar to the level of software architectures. Adaptation steps are considered as abstract model operations working on runtime models. For example, in our previous work~\cite{VogelNHGB10,VG10}, we employed an existing model synchronization engine to maintain an architectural runtime model reflecting a running \emph{Enterprise Java Beans} (EJB) system, and an \emph{Object Constraint Language} (OCL) engine to check architectural constraints on this model. Such engines can be considered as implementations for model operations. On the one hand, this shows that existing model-driven engineering technologies can be seamlessly integrated when using runtime models. On the other hand, the relatively high abstraction level of such model operations makes it feasible to integrate existing implementations of adaptations steps. 

In this regard, the behavior of individual adaptation steps is defined by operations and models, and not by megamodels that are concerned with the interplay of these steps. Thus, megamodels leverage the reuse of implementations for operations and models, and they focus on specifying the interplay and on executing adaptation steps by triggering these implementations in a coordinated and well-defined manner. This motivates the proposed abstraction level of megamodels tackling the flow of operations and models.

In our previous work~\cite{VogelNHGB10,VG10}, we have not modeled the feedback loop or parts of it by megamodels, but presented a conceptual architecture and integrated the different operations by a code-based, static, and specific solution for the loop. However, the examples modeled in this paper with the megamodel language are extended scenarios of the ones used in~\cite{VogelNHGB10,VG10}. This indicates that the proposed modeling language is expressive enough to specify advanced self-adaptive systems with multiple feedback loops (cf.~Sections~\ref{sec:megamodels:multiple-loops}/\ref{sec:megamodels:multiple-loops:hierarchy}).

In contrast to typical models used for developing software, the megamodels created with the proposed modeling language are kept explicit and alive at runtime. Megamodels are executable specifications of feedback loops by defining flows of adaptation steps. Having a megamodel interpreter, the megamodels are directly executed, which provides a flexible solution for feedback loops in contrast to (generated) code-based solutions. This enables adapting a megamodel at runtime, while the interpreter can seamlessly execute the adapted megamodel. Adapting a feedback loop by changing models and the control flow of a megamodel has been discussed in Section~\ref{sec:megamodels:multiple-loops:hierarchy}. Similar adaptations of code-based solutions for feedback loops are likely to be more laborious because there is usually no clear distinction between the code implementing a feedback loop and the code executing a feedback loop. In contrast, explicit megamodels and the interpreter maintain such a distinction.

Finally, keeping megamodels explicit and alive at runtime considerably eases developing hierarchical and adaptive control architectures. A megamodel specifying a feedback loop directly serves as a reflection model that is used to adapt the feedback loop. Thus, there is no need to integrate specific sensors or effectors for monitoring or changing a megamodel resp. the corresponding feedback loop. In contrast, a code-based solution for a feedback loop might require huge efforts to develop sensors and effectors that create and maintain a reflective view on the adaptable feedback loop.

\section{Related Work}
\label{sec:related-work}

In the following, we discuss related approaches to self-adaptive systems, especially to modeling feedback loops.

A popular way to engineer self-adaptive systems are framework-based approaches (cf.~\cite{Salehie&Tahvildari2009}). One example is \emph{Rainbow}~\cite{GarCHSS04} that focuses on reducing development efforts by providing reusable elements of the adaptation engine realizing a feedback loop. The resulting structure of the adaptation steps and thus the feedback loop are rather static and pre-defined by the framework. In general, Rainbow seems to support only one layer that contains one feedback loop.

Another example for a framework is presented in~\cite{SchmWhiGok07}, which proposes a modeling language to specify autonomic systems including the adaptation logic as mappings of assertions to adaptations. The created models do not make the feedback loop explicit and they are only used for generating partial code, but they are not kept alive at runtime to dynamically adjust the adaptation logic.

To summarize, framework-based approaches primarily focus on reducing efforts for developing single feedback loops. The frameworks rather prescribe the structure of the adaptation steps and feedback loops, and they provide no explicit support for adjusting this structure at runtime.

In contrast, we proposed a modeling language to explicitly specify complete feedback loops at an abstraction level similar to software architectures. The created megamodels are kept alive at runtime and they are executed by an interpreter. Thereby, our approach goes one step further than frameworks because it does not prescribe any structure of the adaptation steps or limit the number of feedback loops. Additionally, it supports adapting megamodels at runtime to dynamically adjust the adaptation logic by hierarchically composing feedback loops in layered architectures. 

Concerning the explicit modeling of feedback loops in self-adaptive software systems, several approaches exist.
In~\cite{HGB2010} a UML profile is proposed to make feedback loops and the interplay of multiple feedback loops explicit in architectural design and analysis using UML models. A feedback loop is modeled at the abstraction level of the complete adaptation logic. In contrast, our modeling language additionally supports the specification of individual steps within the adaptation logic that realizes a feedback loop. Moreover, our language and models are used for designing as well as for executing feedback loops at runtime.

A formal reference model to describe self-adaptive systems is presented in~\cite{1809078}. Similar to our language, it supports the description of feedback loops including adaptation steps and models used by the steps. The goal of the reference model is to support the systematic engineering of self-adaptive systems by providing a means to formally describe and evaluate design alternatives. Unlike to our approach, the models are only used in the design phase but not at runtime. 

In~\cite{Vromant+2011}, the issue of interacting adaptation steps as well as interacting feedback loops is addressed by proposing an implementation framework. This framework provides components to address such interactions in order to ease the development. In our approach, we tackle such interactions by explicitly modeling them as model operations coordinating each other by using the same runtime models.

Finally, our modeling language shares characteristics with \emph{UML Activities}~\cite{UML-Superstructure241}. Both languages are similar with respect to modeling flows of actions (in UML) or operations (in our language). However, in contrast to our language, UML does not provide megamodel concepts as first class entities, like a model being itself an element in another model.

\section{Conclusion and Future Work}
\label{sec:conclusion+future-work}

In this paper, we presented a domain-specific modeling language for megamodels to specify and execute feedback loops. The modeling language supports the specification of individual adaptation steps as well as complete feedback loops by modularizing megamodels. Moreover, the interplay between multiple feedback loops is captured. Therefore, feedback loops and adaptation steps are modeled at a high-level of abstraction by means of a flow of model operations working on runtime models. Altogether, feedback loops are made explicit in the megamodels and the development of the adaptation logic realizing a feedback loop is eased.

Moreover, the megamodels are kept alive at runtime and they are executed by an interpreter. This provides the required flexibility to cope with changes of the megamodels at runtime when dynamically adjusting the adaptation logic. This is necessary when feedback loops are composed in a hierarchy to realize adaptive control architectures. Megamodels simplify the hierarchical composition of feedback loops since a megamodel does not only specify a feedback loop but it also serves as the reflection model for any higher level loop. This avoids the need to develop specific sensors and effectors to obtain reflective views on feedback loops.

As future work, we plan to elaborate the modeling language. 
Smoothly integrating formal interface definitions for models and model operations makes the modeling and composition of megamodels more systematic and it enables avenues for analysis. Concerning the metamodel implementation, we are looking into technologies to support the proposed profile.
Moreover, we will discuss the current restrictions on the execution semantics as used by the interpreter. Especially, concurrency will be investigated, which is required to cover continuously running adaptation steps, like monitoring. Finally, we want to integrate the presented megamodel solution with our previous work~\cite{VogelNHGB10,VG10}.

\newcommand{\bibspace}{\vspace{-2mm}}

\end{document}